\documentclass[aps,prd,showpacs,twocolumn,groupedaddress]{revtex4}

\usepackage{amsmath}
\usepackage{amssymb}
\usepackage{graphicx}

\DeclareMathOperator{\arctanh}{arctanh}

\begin{document}

\title{Singularity avoidance and time in quantum gravity}

\author{Andreas Kreienbuehl}
\email[]{k7cw3@unb.ca}
\affiliation{University of New Brunswick\\
Department of Mathematics and Statistics\\
Fredericton, NB E3B 5A3, Canada}

\date{\today}

\begin{abstract}
We consider the quantization of a Friedmann-Robertson-Walker universe. We derive a reduced square-root Hamiltonian by choosing the scale factor as time variable and quantize the theory using Pauli matrices \textit{\`a la} Dirac. From the resulting spinor equation we show that there is no semiclassical wave packet that avoids the big bang singularity. Our work raises the question concerning the relationship between the choice of time and singularity avoidance. 
\end{abstract}


\maketitle

\section{Introduction}
To understand space and time at the Planck scale we have to quantize general relativity (GR). An attempt to achieve this goal consists in the program of canonical quantum gravity. Here, one has to capture the dynamics of a globally hyperbolic spacetime in terms of a Hamiltonian formulation of GR. Following the work of Pirani and others~\cite{pirani:50,pirani:52}, Bergmann and others~\cite{bergmann:50}, as well as Dirac~\cite{dirac:58}, Arnowitt, Deser, and Misner~\cite{arnowitt:62} wrote down the sought-after formulation in terms of metric variables in 1962. Using this result, the quantization can be performed according to Dirac's work on constrained Hamiltonian systems~\cite{dirac:74}, which gives rise to what is called the Wheeler-DeWitt equation~\cite{wheeler:68,dewitt:69}. Another method is referred to as reduced phase space quantization. This procedure has been extensively developed in work by Kucha$\check{\textrm{r}}$~\cite{kuchar:72,kuchar:73} in the 1970s. The idea is to solve the Hamiltonian and diffeomorphism constraints on the classical level. This brings them into a form that, after a Dirac quantization, gives rise to a functional Schr\"odinger equation, called a Tomonaga-Schwinger equation~\cite{tomonaga:46,schwinger:48}. Finally, the program of loop quantum gravity (LQG) provides another way to quantize GR. Here, instead of using metric variables, one works with so called Ashtekar variables~\cite{ashtekar:86}. The origin of this approach lies in work performed by Sen~\cite{sen:82} in 1982.

There are a number of problems that come along with both Dirac and reduced phase space quantization. The most prominent is the one of time~\cite{isham:93,kuchar:92,kiefer:07}. When compared to the Schr\"odinger equation, the Wheeler-DeWitt equation does not have the time derivative on the right-hand side. Among other things, this implies the lack of an obvious Hilbert space to work with. Regarding the reduced phase space quantization, the problem of time seems to be more severe. First of all, the method fails when applied to full GR~\cite{torre:92}. Secondly, when it comes to using the technique for simple models, different classical time choices lead to different reduced Hamiltonians. Therefore the Tomonega-Schwinger equations associated with them are not the same, which raises the question about the validity of one choice of time over any other. Also, as the reduced Hamiltonians are square-root functions and in general time dependent, deriving and solving the functional Schr\"odinger equation is very difficult or even impossible.

In spite of everything, the interest in the ideas and concepts mentioned above is unwaning. This is because the focus has been shifted from the full theory to its applications in simple models. Most prominent are Bianchi models, whereof the isotropic Friedmann-Robertson-Walker (FRW) cosmologies play the most important role. The reason for this is twofold. On the one hand, spatially flat FRW cosmologies provide us with the best model we have for our Universe; on the other, they state that space and time started with a big bang singularity. This is precisely where Einstein's theory of gravity is supposed to break down and where quantum effects are expected to become crucial. In fact, the expectation and hope is that they lead to an avoidance of the big bang. This has been shown to be true not only for FRW cosmologies~\cite{ashtekar:06a,ashtekar:06b,ashtekar:06c,ashtekar:07,husain:03} but also in other cases such as spherically symmetric spacetimes~\cite{husain:08,ashtekar:05,modesto:04,modesto:06}. However, the matter used is as simple as ours, namely a massless scalar field. There is not a single case known to the author where more realistic matter has been successfully incorporated.

In this paper we investigate the problem of the singularity avoidance in a flat FRW universe with a nonnegative cosmological constant and a massless scalar field as the source. In Sec. \ref{s:cl_th} we state the well known classical theory of the model and, in particular, the Hamiltonian formulation thereof. It is in this section where we establish a basis for the quantization. To be more precise, we solve the Hamiltonian constraint by choosing the scale factor of the FRW metric as time variable. This gives rise to a reduced Hamiltonian, which is a square-root function of time and the momentum canonically conjugated to the massless scalar field. In Sec. \ref{s:qu_th} we first perform the quantization with the help of a trick that Dirac used to derive his famous equation from the dispersion relation in special relativity. Then we solve the resulting Schr\"odinger equation; we construct wave packets that are semiclassical at late time and show that they do not avoid the big bang singularity. Finally in Sec. \ref{s:discuss} we give a general discussion and make a comparison to results from loop quantum cosmology (LQC), where a singularity avoidance is achieved.

\section{\label{s:cl_th}Classical Theory}
The model we are investigating can be characterized as follows: First, we use a spatially flat FRW metric
\begin{equation*}
ds^2=-N(t)^2\mathrm{d}t^2+a(t)^2\delta_{ij}\mathrm{d}x^i\mathrm{d}x^j
\end{equation*}
with a nonnegative lapse function $N$ and a scale factor $a$. Second, the source of our spacetime is a massless scalar field $\phi$, minimally coupled to the gravitational field. We choose it to be a function of $t$ only so that the symmetries of the metric are respected. Finally, we have a nonnegative cosmological constant $\Lambda$.

\subsection{Lagrangian formulation}
The field equations and their solutions can be derived from a Lagrangian formulation. One of its main ingredients is the Ricci scalar
\begin{equation*}
R=6\left[\frac{\ddot{a}}{N^2a}+\left(\frac{\dot{a}}{Na}\right)^2-\frac{\dot{N}\dot{a}}{N^3a}\right].
\end{equation*}
If, in addition to the lapse function, we restrict the scale factor to be nonnegative as well, drop an overall factor $V$ representing what is called the fiducial volume and set $\kappa=16\pi G$, we can write
\begin{equation}\label{e:lagrangian}
L=-\frac{6}{\kappa}\left(\frac{a\dot{a}^2}{N}+\frac{\Lambda}{3}Na^3\right)+\frac{a^3\dot{\phi}^2}{2N}
\end{equation}
for the Lagrangian. On the matter side, this gives rise to an energy density and a pressure given by
\begin{equation*}
-T^0_{~0}=\frac{\dot{\phi}^2}{2N^2}~~\mathrm{and}~~T^i_{~i}=\frac{\dot{\phi}^2}{2N^2}
\end{equation*}
respectively. As far as the field equations are concerned we can write down two independent ones for the three unknowns $N$, $a$, and $\phi$ so that a choice has to be made. Since the velocity $\dot{N}$ is absent in the Lagrangian, it is most natural to determine the lapse function. We fix a time variable, a clock. If we set it equal to $1$ then we are dealing with the proper time $\tau$ and, according to~\cite{stephani:03}, the solutions to the field equations are
\begin{equation}\label{e:lambda_zero_sol}
\begin{split}
a(\tau)&=\left(\sqrt{3\lambda}\tau\right)^{1/3},\\
\phi(\tau)&=\phi_{\circ}+\frac{2}{\sqrt{3\kappa}}\ln\left(\sqrt{\lambda}\tau\right)
\end{split}
\end{equation}
for $\Lambda=0$, and
\begin{equation}\label{e:lambda_gr_zero_sol}
\begin{split}
a(\tau)&=\left[\sqrt{\frac{\lambda}{\Lambda}}\sinh\left(\sqrt{3\Lambda}\tau\right)\right]^{1/3},\\
\phi(\tau)&=\phi_{\circ}+\frac{4}{\sqrt{3\kappa}}\arctanh\left(\mathrm{e}^{-\sqrt{3\Lambda}\tau}\right)
\end{split}
\end{equation}
for $\Lambda>0$. The constant $\lambda=\kappa a^6\dot{\phi}^2/4$ is positive and $\phi_{\circ}$ is specified by the initial data. The solutions we presented just now imply a big bang singularity for all $\Lambda\ge0$ (this holds true even for $\Lambda<0$). It does not vanish for other choices of the lapse function as we can simply rescale the time axis. In other words, GR predicts a singularity for the model we study, independent of the lapse function and the cosmological constant.

\subsection{Hamiltonian formulation}
We are interested in studying a quantized version of the given model. The path we choose to get there leads over a Hamiltonian formulation. We initiate it with the specification of the Arnowitt-Deser-Misner (ADM) action~\cite{arnowitt:62}. If we parametrize the metric $h_{ij}$ on spatial hypersurfaces according to $h_{ij}=a^2\delta_{ij}$, where $a$ is any sufficiently differentiable function of $t$, and if we Legendre transform the matter part of the Lagrangian in Eq.\ (\ref{e:lagrangian}), the ADM action in canonical form is given by
\begin{equation*}
S=\int(p_a\dot{a}+p_{\phi}\dot{\phi}-H)~\mathrm{d}t.
\end{equation*}
The momenta are
\begin{equation}\label{e:momenta}
p_a=-\frac{12V}{\kappa}\frac{|a|\dot{a}}{N},~~p_{\phi}=V\frac{|a|^3\dot{\phi}}{N},
\end{equation}
the nonvanishing Poisson brackets $\left\{p_a,a\right\}=1$, $\left\{p_{\phi},\phi\right\}=1$, and the Hamiltonian
\begin{equation*}
H=N\left(-\frac{\kappa}{24V}\frac{p_a^2}{|a|}+\frac{2V\Lambda}{\kappa}|a|^3+\frac{1}{2V}\frac{p_{\phi}^2}{|a|^3}\right)=0
\end{equation*}
is constrained to vanish. We include the fiducial volume $V$ in order to keep track of all units. On the Lagrangian level, the vanishing of $H$ is equivalent to the Friedmann equation. However, on the Hamiltonian level it implies a static universe and to get dynamics, a time variable has to be chosen. We focus on a clock defined by $\sqrt{\lambda}t=a(t)$ which, suppressing the argument $p_{\phi}$, gives rise to the reduced Hamiltonian
\begin{equation}\label{e:reduced_hamiltonian}
h_a(t)=\left(\frac{48V^2\lambda^3\Lambda}{\kappa^2}t^4+\frac{12}{\kappa}\frac{p_{\phi}^2}{t^2}\right)^{1/2}.
\end{equation}
Since we want the choice for the time variable to be preserved as $t$ changes, we have to have $1=\{H,\lambda^{-1/2}a\}$ which, using Eq.\ (\ref{e:reduced_hamiltonian}), implies
\begin{equation*}
N(t)=\left(\frac{\Lambda}{3}t^2+\frac{\kappa}{12V^2\lambda^3}\frac{p_{\phi}(t)^2}{t^4}\right)^{-1/2}.
\end{equation*}
Had we decided to go with the negative square-root of the reduced Hamiltonian, we would refer to a contracting phase of the Universe (with respect to any coordinate time that is monotonically increasing). This can be seen from the definition of $p_a$ in Eq.\ (\ref{e:momenta}). We stick with the positive square-root since the clock we defined describes an expanding phase. Now, for $t>0$, Hamilton's equations of motion are solved by
\begin{equation}\label{e:cl_sol_lam_zero}
\phi(t)=\phi_{\circ}+\sqrt{\frac{12}{\kappa}}\ln\left(\sqrt{\lambda}t\right)
\end{equation}
for $\Lambda=0$, and
\begin{equation}\label{e:cl_sol_lam_gr_zero}
\phi(t)=\phi_{\circ}+\frac{2}{\sqrt{3\kappa}}\ln\left[\left(\frac{\Lambda}{\lambda}+\frac{1}{\lambda^3t^6}\right)^{1/2}-\frac{1}{\lambda^{3/2}t^3}\right]
\end{equation}
for $\Lambda>0$. In both cases, $p_{\phi}(t)\doteq p_{\phi}$ is a positive constant. If it is given by $2V(\lambda/\kappa)^{1/2}$ we can recover Eqs.\ (\ref{e:lambda_zero_sol}) and (\ref{e:lambda_gr_zero_sol}) by plugging in the proper time variable $\tau$, which can be constructed from the lapse function. As can be seen from Eq.\ (\ref{e:lambda_gr_zero_sol}), this means that if $\Lambda>0$, the expansion of the Universe is accelerating at late time. Finally, the equations
\begin{align*}
R(t)&=4\Lambda-\frac{\kappa}{2V^2\lambda^3}\frac{p_{\phi}^2}{t^6},\\
-T^0_{~0}(t)&=\frac{1}{2V^2\lambda^3}\frac{p_{\phi}^2}{t^6},~~\mathrm{and}~~T^i_{~i}(t)=\frac{1}{2V^2\lambda^3}\frac{p_{\phi}^2}{t^6}
\end{align*}
imply that we have the same big bang singularity as in the Lagrangian formulation.

\section{\label{s:qu_th}Quantum Theory}
Because of the square-root, the Hamiltonian in Eq.\ (\ref{e:reduced_hamiltonian}) is troublesome. Even though we can insert a Schr\"odinger representation for the momentum $p_{\phi}$ and, at least for $\Lambda\geq0$, directly get a self-adjoint and positive square-root Hamiltonian
\begin{equation}\label{e:square_root_ham}
\hat{h}_a(t)=\left(\frac{48V^2\lambda^3\Lambda}{\kappa^2}t^4-\frac{12}{\kappa}\frac{\hbar^2}{t^2}\partial_{\phi}^2\right)^{1/2}
\end{equation}
on $L^2(\mathbb{R})$, we cannot solve the time dependent Schr\"o\-ding\-er equation. As has been carefully pointed out in~\cite{blyth:75}, an attempt to circumvent this problem by solving the Klein-Gordon equation fails because of the explicit time dependence of the Hamiltonian. All we can do with the square-root Hamiltonian is to solve the time independent Schr\"odinger equation
\begin{equation*}
\hat{h}_a(t)\psi_{E(t)}(t,\phi)=E(t)\psi_{E(t)}(t,\phi).
\end{equation*}
The reason for this is that, here, we can square on both sides. It leads to the eigenstate
\begin{equation}\label{e:eigenstate}
\psi_{E(t)}(t,\phi)=\psi_{\scriptscriptstyle-}(t)\mathrm{e}^{-i\mathcal{P}(t)\phi/\hbar}+\psi_{\scriptscriptstyle+}(t)\mathrm{e}^{i\mathcal{P}(t)\phi/\hbar},
\end{equation}
where the momentum $\mathcal{P}(t)$ is given by
\begin{equation}\label{e:momentum}
\mathcal{P}(t)=\left(\frac{\kappa}{12}E(t)^2t^2-\frac{4V^2\lambda^3\Lambda}{\kappa}t^6\right)^{1/2},
\end{equation}
and the functions $\psi_{\scriptscriptstyle\pm}(t)$ are determined by the initial data. Note that the expression for $\mathcal{P}(t)$ introduces a lower bound on $E(t)$, specified by the cosmological constant. Now, because of
\begin{equation}\label{e:commutator}
\left[\hat{h}_a(t_l),\hat{h}_a(t)\right]=0,
\end{equation}
the eigenstate in Eq.\ (\ref{e:eigenstate}) is an eigenstate at a later time $t_l\geq t$ as well. Therefore, if the Universe is in an eigenstate at $t_l$, its energy $E(t_l)$ is given by
\begin{equation}\label{e:energy_relation}
E(t_l)^2t_l^2=E(t)^2t^2+\frac{48V^2\lambda^3\Lambda}{\kappa^2}(t_l^6-t^6).
\end{equation}
This is a relation between the vacuum energies $E(t_l)$ and $E(t)$. The difference between them comes from the energy introduced by the cosmological constant. Using this relation and Eq.\ (\ref{e:commutator}), we can evolve the eigenstate in Eq.\ (\ref{e:eigenstate}) to $t_l$. For $t>0$ and $\Lambda>0$ this yields
\begin{equation}\label{e:eigen_evol}
\psi_{E(t_l)}(t_l,\phi)=\mathrm{e}^{-i\int_{t}^{t_l}E(t_l')~\mathrm{d}t_l'/\hbar}\psi_{E(t)}(t,\phi),
\end{equation}
where the integral in the exponent is 
\begin{align*}
&\int_{t}^{t_l}E(t_l')~\mathrm{d}t_l'=\frac{E(t_l)t_l}{3}-\frac{E(t)t}{3}\\
&-\frac{2}{\sqrt{3\kappa}}\mathcal{P}(t)\ln\left[\left(\frac{t}{t_l}\right)^3\frac{\sqrt{12/\kappa}\mathcal{P}(t)+E(t_l)t_l}{\sqrt{12/\kappa}\mathcal{P}(t)+E(t)t}\right].
\end{align*}
The case $\Lambda=0$ is discussed in~\cite{blyth:75}. This is all we can say about the solutions to the time independent Schr\"odinger equation for the square-root Hamiltonian. Again, because of the square-root, the time dependent Schr\"odinger equation cannot be solved. Fortunately this is not the end of the story as we can quantize the reduced Hamiltonian in another way.

\subsection{\label{ss:dirac}Dirac's trick}
Motivated by Dirac's analysis of the Klein-Gordon equation, which led him to derive the Dirac equation, we start with an Ansatz for the Hamiltonian operator of the form
\begin{equation}\label{e:ansatz}
\hat{h}_a(t)=A\left(\frac{48V^2\Lambda\lambda^3}{\kappa^2}\right)^{1/2}t^2+B\left(\frac{12}{\kappa}\right)^{1/2}\frac{\hat{p}_{\phi}}{t},
\end{equation}
where $A$ and $B$ have to be Hermitian matrices. Note that we set $t>0$. If we square the Ansatz on both sides and make a comparison with the squared  Hamiltonian from Eq.\ (\ref{e:square_root_ham}), we get a system of equations for $A$ and $B$ that is solved by any pair of distinct Pauli matrices and transformations by an arbitrary nonsingular matrix thereof. Here, we set
\begin{equation*}
A=\begin{bmatrix}0&1\\1&0\end{bmatrix}~~\mathrm{and}~~B=\begin{bmatrix}1&0\\0&-1\end{bmatrix}.
\end{equation*}
The motivation for this choice will become clear further down. Using a Schr\"odinger representation for $\hat{p}_{\phi}$ the Hamiltonian is now
\begin{equation*}
\hat{h}_a(t)=\begin{bmatrix}-\beta\frac{i\hbar}{t}\partial_{\phi}&\alpha t^2\\\alpha t^2&\beta\frac{i\hbar}{t}\partial_{\phi}\end{bmatrix},
\end{equation*}
where $\alpha$ and $\beta$ are chosen in accordance with Eq.\ (\ref{e:ansatz}). It is now clear that $\Lambda<0$ is not possible as it prevents the Hamiltonian from being Hermitian. If we define $\mathcal{H}=L^2(\mathbb{R})\oplus L^2(\mathbb{R})$ as the Hilbert space, with the inner product of two states $\Psi,\Omega\in\mathcal{H}$ given by
\begin{equation*}
\langle\Psi|\Omega\rangle=\int\Psi^{\dag}\Omega~\mathrm{d}\phi/\beta,
\end{equation*}
we make $\hat{h}_a(t)$ a self-adjoint operator on $\mathcal{H}$. Also, the induced norm of a state $\Psi=(\psi,\omega)^{\mathrm{t}}\in\mathcal{H}$ is conserved. This can be seen from the continuity equation
\begin{equation*}
\partial_t\left(|\psi|^2+|\omega|^2\right)+\partial_{\phi}\left[\beta/t\left(|\psi|^2-|\omega|^2\right)\right]=0,
\end{equation*}
whose origin lies in the time dependent Schr\"odinger equation
\begin{equation}\label{e:schr_equ}
\hat{h}_a(t)\Psi(t,\phi)=i\hbar\partial_t\Psi(t,\phi).
\end{equation}
Note that the components $\psi$ and $\omega$ decouple in this equation if we neglect the cosmological constant. The same decoupling occurs when $t$ approaches zero. This is the limit in which the big bang singularity can be found on the classical level and it motivates our choice for the matrices $A$ and $B$. 

The question remains whether the singularity can be avoided when we use the quantum mechanical formulation we built just now. To answer it we first have to solve the time dependent Schr\"odinger equation. Then we have to construct semiclassical wave packets and check whether the expectation values of the Ricciscalar, the energy density, and the pressure, with respect to these states, remain finite when $t$ goes to zero. In the appendix  we consider other choices for the matrices $A$ and $B$ and we show that the final conclusions remain the same.

Before we go into the discussion on the singularity avoidance, we take a look at the time independent Schr\"odinger equation
\begin{equation*}
\hat{h}_a(t)\Psi_{E(t)}(t,\phi)=E(t)\Psi_{E(t)}(t,\phi).
\end{equation*}
The solutions are
\begin{align}\label{e:square_eigenstate}
\Psi_{E(t)}(t,\phi)=&\left(\begin{array}{c}0\\1\end{array}\right)\psi_{\scriptscriptstyle-}(t)\mathrm{e}^{-i\frac{E(t)t}{\beta\hbar}\phi}\nonumber\\
&+\left(\begin{array}{c}1\\0\end{array}\right)\psi_{\scriptscriptstyle+}(t)\mathrm{e}^{i\frac{E(t)t}{\beta\hbar}\phi}
\end{align}
for $\Lambda=0$, and
\begin{equation*}
\begin{split}
\Psi_{E(t)}(t,\phi)=&\left(\begin{array}{c}1\\\frac{E(t)}{\alpha t^2}+\frac{\beta}{\alpha t^3}\mathcal{P}(t)\end{array}\right)\psi_{\scriptscriptstyle-}(t)\mathrm{e}^{-i\mathcal{P}(t)\phi/\hbar}\\
&+\left(\begin{array}{c}1\\\frac{E(t)}{\alpha t^2}-\frac{\beta}{\alpha t^3}\mathcal{P}(t)\end{array}\right)\psi_{\scriptscriptstyle+}(t)\mathrm{e}^{i\mathcal{P}(t)\phi/\hbar}
\end{split}
\end{equation*}
for $\Lambda>0$. As before, the functions $\psi_{\scriptscriptstyle\pm}(t)$ are specified by the initial data and the momentum $\mathcal{P}(t)$ is defined in Eq.\ (\ref{e:momentum}). Here, $\mathcal{P}(t)$ introduces both an upper bound on the negative energies and a lower bound on the positive energies by means of the cosmological constant. Using the same reasoning as for Eqs.\ (\ref{e:energy_relation}) and (\ref{e:eigen_evol}), if $\Lambda=0$ the energy at $t_l\geq t$ is given by $E(t_l)=E(t)t/t_l$ and the eigenstate in Eq.\ (\ref{e:square_eigenstate}) by
\begin{equation*}
\Psi_{E(t_l)}(t_l,\phi)=\left(\frac{t}{t_l}\right)^{iE(t)t/\hbar}\Psi_{E(t)}(t,\phi).
\end{equation*}
However, if $\Lambda>0$, we cannot write down similar expressions for the energy and the eigenstate at $t_l$ because we now have
\begin{equation*}
\left[\hat{h}_a(t_l),\hat{h}_a(t)\right]\neq0.
\end{equation*}
This implies that $\hat{h}_a(t_l)$ and $\hat{h}_a(t)$ do not have common eigenstates, a property which is used for the computation of $E(t_l)$. Also, it means that the time ordering symbol in the Dyson series, which defines the evolution of the eigenstates, cannot be neglected. This being said, we conclude the discussion of the time independent Schr\"odinger equation.

\subsection{Singularity avoidance?}

\begin{figure}[ht!]
\includegraphics[width=\columnwidth]{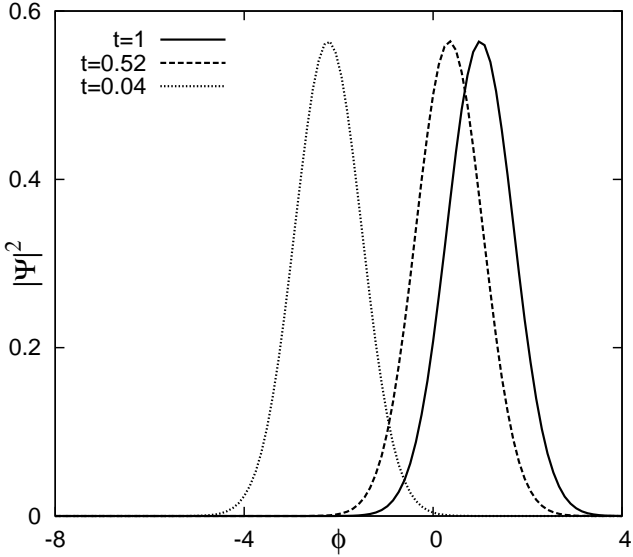}
\caption{\label{f:lambda_zero} We chose $\phi_{\circ}=1$, $k_{\circ}=4$, and $W=0.5$. The peak traces the classical solution and evolves freely.}
\end{figure}

\paragraph{Case $\Lambda=0$.}
We need to solve the time dependent Schr\"odinger equation. This is a task that can be simplified if we replace $t$ and $\phi$ by the dimensionless coordinates $\sqrt{\lambda}t$ and $\phi/\beta$ respectively. Namely, it allows us to rewrite Eq.\ (\ref{e:schr_equ}) and get
\begin{equation}\label{e:schr}
\begin{bmatrix}t\partial_t+\partial_{\phi}&0\\0&t\partial_t-\partial_{\phi}\end{bmatrix}\Psi(t,\phi)=0,
\end{equation}
for which the solution is given by
\begin{equation*}
\psi(t,\phi)=f(\phi-\ln(t))~~\mathrm{and}~~\omega(t,\phi)=g(\phi+\ln(t)).
\end{equation*}
The functions $f$ and $g$ are in $L^2(\mathbb{R})$, but otherwise arbitrary. Therefore, we can write down a wave packet
\begin{equation*}
\Psi(t,\phi)=m\left(\begin{array}{c}\mathrm{e}^{-\frac{(\phi-\phi_{\circ}-\ln(t))^2}{2W^2}}\mathrm{e}^{ik_{\circ}(\phi-\phi_{\circ}-\ln(t))}\\
0\end{array}\right)
\end{equation*}
with normalization constant $m=\pi^{-1/4}W^{-1/2}$. As can be seen from
\begin{equation}\label{e:exp_momentum}
\langle\hat{p}_{\phi}\rangle_{\Psi}=\hbar k_{\circ}~~\mathrm{and}~~\langle\hat{p}_{\phi}^2\rangle_{\Psi}=\hbar^2k_{\circ}^2+\frac{\hbar^2}{2W^2},
\end{equation}
among the constants $k_{\circ}$ and $W$ the former describes the momentum of the wave packet, whereas the latter characterizes the quantum fluctuations of the squared momentum operator. Since the expectation values
\begin{equation*}
\langle\hat{\phi}\rangle_{\Psi}=\phi_{\circ}+\ln(t)~~\mathrm{and}~~\langle\hat{\phi}^2\rangle_{\Psi}=\left(\phi_{\circ}+\ln(t)\right)^2+\frac{W^2}{2}
\end{equation*}
imply that $\Psi$ is always peaked over the classical solution from Eq.\ (\ref{e:cl_sol_lam_zero}), with standard deviation $W/\sqrt{2}$, we have a semiclassical wave packet if we require $k_{\circ}\gg10^{-34}$ (the numerical value of $\hbar$) and $1\gg W\gg10^{-34}$ (in units where $\beta\doteq1$). Now, the point is that this holds true for all $t>0$. The evolution has no impact on the form of the wave packet and tells us only where in the $(t,\phi)$-plane $\Psi$ is most likely to be found. In Fig.\ \ref{f:lambda_zero} we plot the probability density function $|\Psi|^2$ for small and decreasing values of $t$.
It indicates that the evolution is undisturbed, and that the wave packet follows the classical solution. 

As far as the big bang singularity is concerned, the equation 
\begin{equation*}
\lim_{t\to0}|\Psi(t,\phi)|^2=0
\end{equation*}
implies that the probability density function vanishes if $\phi\in\mathbb{R}$ is fixed and $t$ approaches zero. However, this only reflects the fact that the wave packet bypasses finite values of $\phi$ and follows the classical, diverging solution given in Eq.\ (\ref{e:cl_sol_lam_zero}). Indeed, from Eq.\ (\ref{e:exp_momentum}) we get
\begin{equation}\label{e:quantum_big_bang}
\begin{split}
\langle\hat{R}(t)\rangle_{\Psi}&=-\frac{\kappa}{2V^2\lambda^3}\frac{\langle\hat{p}_{\phi}^2\rangle_{\Psi}}{t^6}\xrightarrow{t\to0}-\infty,\\
-\langle\hat{T}^0_{~0}(t)\rangle_{\Psi}&=\frac{1}{2V^2\lambda^3}\frac{\langle\hat{p}_{\phi}^2\rangle_{\Psi}}{t^6}\xrightarrow{t\to0}\infty
\end{split}
\end{equation}
and
\begin{equation}\label{e:q_b_b_part2}
\langle\hat{T}^i_{~i}(t)\rangle_{\Psi}=\frac{1}{2V^2\lambda^3}\frac{\langle\hat{p}_{\phi}^2\rangle_{\Psi}}{t^6}\xrightarrow{t\to0}\infty.
\end{equation}
The big bang singularity is not avoided.

The above conclusion holds true for arbitrary wave packets. Any $\Psi$ that solves Eq.\ (\ref{e:schr}) can only have an upper component if it is to be semiclassical. Since this component is a function of $\phi-\phi_{\circ}-\ln(t)$, an integration over $\phi$ can be replaced by one over $\phi-\phi_{\circ}-\ln(t)$, thereby removing the $t$-dependence. Therefore, the expectation value $\langle\hat{p}_{\phi}^2\rangle_{\Psi}$ is independent of $t$. Since the squared momentum operator has a spectrum on $L^2(\mathbb{R})$ that is strictly positive (except for the irrelevant constant functions), the big bang singularity cannot be avoided.

\begin{figure}[ht!]
\includegraphics[width=\columnwidth]{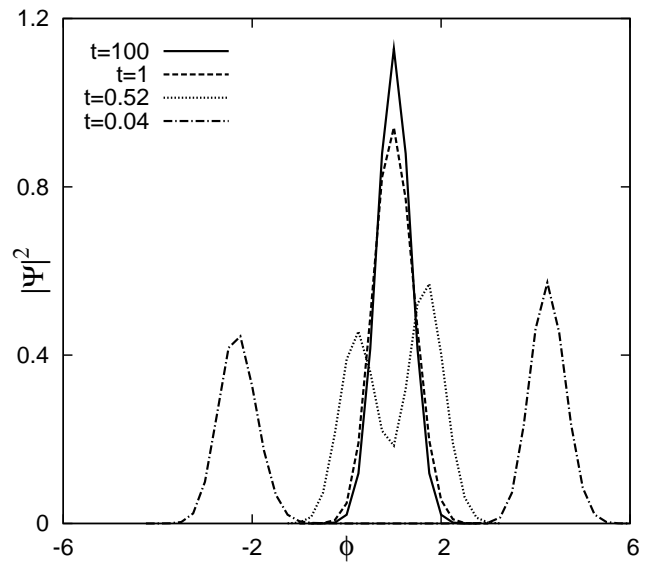}
\caption{\label{f:lambda_gr_zero} We chose $\Lambda=\lambda$, $\phi_{\circ}=1$, $k_{\circ}=4$, and $W=0.5$. Starting at $\phi=-6$, we computed $|\Psi|^2$ after every $\triangle\phi=0.25$. The wave packet is no longer semiclassical for small $t$.}
\end{figure}

\paragraph{Case $\Lambda>0$.}
Here, we work with the dimensionless coordinates $(\alpha/(6\hbar))^{1/3}t$ and $\phi/\beta$ so that the Schr\"odinger equation becomes
\begin{equation*}
\begin{bmatrix}t\partial_t+\partial_{\phi}&i6t^3\\i6t^3&t\partial_t-\partial_{\phi}\end{bmatrix}\Psi(t,\phi)=0.
\end{equation*}
Its solution is given by
\begin{equation*}
\psi(t,\phi)=it^{3/2}\int\big[j(k)J_{\nu}(2t^3)+y(k)Y_{\nu}(2t^3)\big]\textrm{e}^{ik\phi}~\mathrm{d}k
\end{equation*}
and
\begin{equation*}
\omega(t,\phi)=t^{3/2}\int\big[j(k)J_{\nu+1}(2t^3)+y(k)Y_{\nu+1}(2t^3)\big]\textrm{e}^{ik\phi}~\mathrm{d}k,
\end{equation*}
where $\nu=-1/2-ik/3$. The functions $J_{\nu}$ and $Y_{\nu}$ are the Bessel functions of the first and the second kind of order $\nu$ respectively. See~\cite{abramowitz:72} and~\cite{gradshteyn:07} for useful details related to these functions. If we set
\begin{equation*}
j(k)=n\cosh\left(\frac{\pi}{6}k\right)\mathrm{e}^{-\frac{W^2}{2}(k-k_{\circ})^2}\mathrm{e}^{-ik\left(\phi_{\circ}+\frac{1}{6}\ln(\Lambda/\lambda)\right)}
\end{equation*}
and
\begin{equation*}
y(k)=in\sinh\left(\frac{\pi}{6}k\right)\mathrm{e}^{-\frac{W^2}{2}(k-k_{\circ})^2}\mathrm{e}^{-ik\left(\phi_{\circ}+\frac{1}{6}\ln(\Lambda/\lambda)\right)}
\end{equation*}
where $n=2^{1/2}\pi^{-1/4}W^{-1/2}$ is a normalization constant, we get the wave packet
\begin{align}\label{e:full_wave_packet}
\Psi(t,\phi)=&~nt^{3/2}\int\bigg[\cosh\left(\frac{\pi}{6}k\right)\bigg(\begin{array}{c}iJ_{\nu}(2t^3)\\J_{\nu+1}(2t^3)\end{array}\bigg)\nonumber\\
&+\sinh\left(\frac{\pi}{6}k\right)\bigg(\begin{array}{c}-Y_{\nu}(2t^3)\\iY_{\nu+1}(2t^3)\end{array}\bigg)\bigg]\nonumber\\
&\cdot\mathrm{e}^{-\frac{W^2}{2}(k-k_{\circ})^2}\mathrm{e}^{ik\left(\phi-\phi_{\circ}-\frac{1}{6}\ln(\Lambda/\lambda)\right)}~\mathrm{d}k.
\end{align}
The choice for the functions $j$ and $y$ is motivated by the fact that when $t$ is large, $\Psi$ is semiclassical. Namely, in this limit we have
\begin{align}\label{e:late_t_wave}
\Psi(t,\phi)\sim&~m\left(\begin{array}{c}i\cos(4t^3)\\
\sin(4t^3)\end{array}\right)\nonumber\\
&\cdot\mathrm{e}^{-\frac{\left(\phi-\phi_{\circ}-\frac{1}{6}\ln(\Lambda/\lambda)\right)^2}{2W^2}}\mathrm{e}^{ik_{\circ}\left(\phi-\phi_{\circ}-\frac{1}{6}\ln(\Lambda/\lambda)\right)}.
\end{align}
Because of our choice of $n$, this wave packet is normalized and we can compute the expectation values
\begin{equation}\label{e:exp_varphi}
\langle\hat{\phi}\rangle_{\Psi}\sim\phi_{\circ}+\frac{1}{6}\ln\left(\frac{\Lambda}{\lambda}\right)
\end{equation}
and
\begin{equation*}
\langle\hat{\phi}^2\rangle_{\Psi}\sim\left[\phi_{\circ}+\frac{1}{6}\ln\left(\frac{\Lambda}{\lambda}\right)\right]^2+\frac{W^2}{2}.
\end{equation*}
A comparison with Eq.\ (\ref{e:cl_sol_lam_gr_zero}) shows that, at late time, $\Psi$ is peaked over the classical solution. Also, Eq.\ (\ref{e:late_t_wave}) gives 
\begin{equation}\label{e:results}
\langle\hat{p}_{\phi}\rangle_{\Psi}\sim\hbar k_{\circ}~~\mathrm{and}~~\langle\hat{p}_{\phi}^2\rangle_{\Psi}\sim\hbar^2k_{\circ}^2+\frac{\hbar^2}{2W^2}
\end{equation}
so that as in the previous case, if we require $k_{\circ}\gg10^{-34}$ (the numerical value of $\hbar$) and $1\gg W\gg10^{-34}$ (in units where $\beta\doteq1$), we have a semiclassical wave packet. However, this does not hold true for all $t>0$, the reason being that $\Psi$ has two components and if $t\ll1$, they move in different directions with respect to the $\phi$-axis. This becomes evident if we expand Eq.\ (\ref{e:full_wave_packet}) about $t=0$, from which we get
\begin{align*}
\psi(t,\phi)\sim i\frac{n}{\pi}\int&\cosh\left(\frac{\pi}{6}k\right)\Gamma(-\nu)\\
&\cdot\mathrm{e}^{-\frac{W^2}{2}(k-k_{\circ})^2}\mathrm{e}^{ik(\phi-\phi_{\circ}-\ln(t))}~dk
\end{align*}
and
\begin{align*}
\omega(t,\phi)\sim-i\frac{n}{\pi}\int&\sinh\left(\frac{\pi}{6}k\right)\Gamma(\nu+1)\\
&\cdot\mathrm{e}^{-\frac{W^2}{2}(k-k_{\circ})^2}\mathrm{e}^{ik(\phi-\phi_{\circ}+\ln(t))}~dk,
\end{align*}
where $\Gamma$ denotes the gamma function. A complete discussion of this function can be found in~\cite{abramowitz:72,gradshteyn:07}. Now, $|\psi|^2$ is a Gaussian function in $\phi-\phi_{\circ}-\ln(t)$, and $|\omega|^2$ is one in $\phi-\phi_{\circ}+\ln(t)$. Therefore, when $t$ gets closer to zero their peaks approach $\phi=-\infty$ and $\phi=\infty$, respectively. As can be seen from Eq.\ (\ref{e:cl_sol_lam_gr_zero}), the component $\psi$ follows the classical solution. However, the component $\omega$ does not and the wave packet is no longer semiclassical. This is illustrated in Fig.\ \ref{f:lambda_gr_zero}, where we plot the full wave packet from Eq.\ (\ref{e:full_wave_packet}) for several values of $t$.

The result in Eq.\ (\ref{e:exp_varphi}) is only true for large $t$. The expectation values in Eq. (\ref{e:results}) however, can be computed using Eq.\ (\ref{e:full_wave_packet}) and are exact for all $t$. Therefore, the Ricci scalar, the energy density and the pressure diverge as $t$ approaches zero, as it happens in Eqs.\ (\ref{e:quantum_big_bang}) and (\ref{e:q_b_b_part2}). In conclusion, the big bang singularity is not avoided. 

The value of the cosmological constant has no effect on the conclusion concerning the singularity avoidance. It simply serves the purpose of giving rise to a time dependent mass that vanishes when $t$ goes to zero. The choice of the wave packet does not matter either. Namely, all $\Psi$ have two components, $\psi$ and $\omega$, whereof for small $t$ the former has to be a function of $\phi-\phi_{\circ}-\ln(t)$ and the latter one of $\phi-\phi_{\circ}+\ln(t)$. Thus, they give rise to an expectation value $\langle\hat{p}_{\phi}^2\rangle_{\Psi}$ that is time independent if $t\ll1$. As it is also strictly positive on $L^2(\mathbb{R})\oplus L^2(\mathbb{R})$, the big bang cannot be avoided.

\section{\label{s:discuss}Discussion}
In the present paper we quantized a flat FRW universe with a nonnegative cosmological constant and a massless scalar field as the source. To do so we first chose the scale factor as time variable, which gave rise to a reduced, time dependent square-root Hamiltonian. Then, in order to be able to perform the quantization, we referred to a famous trick that Dirac used to derive the equation that is named after him. This led to a Schr\"odinger equation that is a system of two coupled linear first-order partial differential equations. We solved it and constructed wave packets that are semiclassical at late time. Next we evolved the wave packets toward the point, where the big bang singularity can be found on the classical level. In case of a vanishing cosmological constant, this led to a quantum big bang as the wave packet remains semiclassical for all time. As far as positive cosmological constants are concerned, the big bang singularity could not be avoided either. This happened, even though the wave packet was no longer semiclassical in the vicinity of the singular point. Finally, we showed that all semiclassical wave packets lead to a quantum big bang.

The clock we chose is expressed in terms of metric variables. It is a good clock as it is closely related to the Hubble time, which is the most natural choice for FRW cosmologies. Blyth and Isham~\cite{blyth:75} considered the same time variable we do. However, their spacetime has a negative curvature and the cosmological constant is set equal to zero. Even though they find a reduced time dependent square-root Hamiltonian, and write down the associated spectrum and eigenstates, no statement is made regarding singularity avoidance. 
 
Clocks are made of matter so that it makes sense to try to use the massless scalar field as a time variable. This is what Blyth and Isham~\cite{blyth:75} do. However, for this calculation they modify our model by letting the spacetime have a nonnegative curvature and neglecting the cosmological constant. They compute the energy spectrum and the eigenstates from the squared, time independent Schr\"odinger equation and show that no wave packet avoids the big bang singularity. 

Another result is found by Ashtekar and his collaborators~\cite{ashtekar:06a,ashtekar:06b,ashtekar:06c,ashtekar:07}. Their model is a FRW universe with a nonnegative curvature, an arbitrary cosmological constant and a massless scalar field as the source. The quantization is performed using techniques from LQC. To be more precise, they use a polymer representation~\cite{corichi:07} for the gravitational part and a Schr\"odinger representation for the matter part of the Hamiltonian constraint. The scalar field is identified as a time variable and a finite difference equation is derived, which leads to a resolution of the big bang singularity. 

It is really a combination of both the clock and the polymer representation that is responsible for the result above. On the one hand, the clock gives rise to a time independent Hamiltonian; on the other, the polymer representation introduces an upper bound on the gravitational kinetic term. This causes the matter part to be bounded too, which leads to the singularity avoidance. 

The choice of the clock is very important in the context of singularity avoidance. In the LQC case~\cite{ashtekar:06a,ashtekar:06b,ashtekar:06c,ashtekar:07}, the scalar field is chosen as time variable. It gives rise to a time independent Hamiltonian and an avoidance of the big bang singularity is found. In our work the scale factor is chosen as clock, the Hamiltonian is time dependent and the singularity is frozen into the model. (We think that a polymer representation of the matter part of the Hamiltonian does not resolve the singularity because replacing $\hat{p}_{\phi}$ by $\sin(\mu p_{\phi})/\mu$ in Eq.\ (\ref{e:ansatz}) does not remove the divergence in the small $t$ limit. However, this polymerization is a conjecture which requires further study to establish.) This also occurs in~\cite{blyth:75} and in~\cite{husain:87}, where the Gowdy universe is considered. In each case, the singularity is unavoidable as soon as the clock is introduced.

What clocks can we choose if the scalar field is massive? Except for the simplest cases, any clock leads to a time dependent Hamiltonian. Also, at least in the context of reduced phase space quantization, the Hamiltonians are square-root functions, and quantizing them is very difficult. Finally, as far as solving the Wheeler-DeWitt equation is concerned, it is not obvious which clocks can be chosen through the WKB approximation~\cite{kiefer:93}.

Altogether, the ultimate question is and remains what can be concluded from minisuperspace models.
\\
\appendix*

\section{Matrix Representations}
First of all, we have to note that the Schr\"odinger equation (\ref{e:schr_equ}) can be written as
\begin{equation*}
\left(-i\hbar\gamma^{\mu}\partial_{\mu}+\alpha'\mathrm{e}^{3t}\right)\Psi(t,\phi)=0.
\end{equation*}
Here we factored out the matrix $A$ and replaced $t$ and $\phi$ by the dimensionless coordinates $\ln(\sqrt{\lambda}t)$ and $\phi/\beta$ respectively, set $\alpha'=\alpha/\lambda^{3/2}$, $\partial_0=\partial_t$ and $\partial_1=\partial_{\phi}$, then used the gamma matrices
\begin{equation*}
\gamma^0=\begin{bmatrix}0&1\\1&0\end{bmatrix}~~\mathrm{and}~~\gamma^1=\begin{bmatrix}0&-1\\1&0\end{bmatrix}.
\end{equation*}
What we wrote down is the Dirac equation in two spacetime dimensions. It follows from its covariance that the results we derived are independent of the choice we made for the matrices $A$ and $B$.
\begin{acknowledgments}
This work was supported in part by the Natural Science and Engineering Research Council of Canada. The author would like to thank S. Seahra and G. M. Hossain for discussions, and V. Husain in particular for suggesting Dirac's trick.
\end{acknowledgments}

\bibliography{d}

\begin{thebibliography}{34}
\expandafter\ifx\csname natexlab\endcsname\relax\def\natexlab#1{#1}\fi
\expandafter\ifx\csname bibnamefont\endcsname\relax
  \def\bibnamefont#1{#1}\fi
\expandafter\ifx\csname bibfnamefont\endcsname\relax
  \def\bibfnamefont#1{#1}\fi
\expandafter\ifx\csname citenamefont\endcsname\relax
  \def\citenamefont#1{#1}\fi
\expandafter\ifx\csname url\endcsname\relax
  \def\url#1{\texttt{#1}}\fi
\expandafter\ifx\csname urlprefix\endcsname\relax\def\urlprefix{URL }\fi
\providecommand{\bibinfo}[2]{#2}
\providecommand{\eprint}[2][]{\url{#2}}

\bibitem[{\citenamefont{Pirani and Schild}(1950)}]{pirani:50}
\bibinfo{author}{\bibfnamefont{F.~A.~E.} \bibnamefont{Pirani}}
  \bibnamefont{and} \bibinfo{author}{\bibfnamefont{A.}~\bibnamefont{Schild}},
  \bibinfo{journal}{Phys. Rev.} \textbf{\bibinfo{volume}{79}},
  \bibinfo{pages}{986} (\bibinfo{year}{1950}).

\bibitem[{\citenamefont{Pirani et~al.}(1952)\citenamefont{Pirani, Schild, and
  Skinner}}]{pirani:52}
\bibinfo{author}{\bibfnamefont{F.~A.~E.} \bibnamefont{Pirani}},
  \bibinfo{author}{\bibfnamefont{A.}~\bibnamefont{Schild}}, \bibnamefont{and}
  \bibinfo{author}{\bibfnamefont{R.}~\bibnamefont{Skinner}},
  \bibinfo{journal}{Phys. Rev.} \textbf{\bibinfo{volume}{87}},
  \bibinfo{pages}{452} (\bibinfo{year}{1952}).

\bibitem[{\citenamefont{Bergmann et~al.}(1950)\citenamefont{Bergmann, Penfield,
  Schiller, and Zatzkis}}]{bergmann:50}
\bibinfo{author}{\bibfnamefont{P.~G.} \bibnamefont{Bergmann}},
  \bibinfo{author}{\bibfnamefont{R.}~\bibnamefont{Penfield}},
  \bibinfo{author}{\bibfnamefont{R.}~\bibnamefont{Schiller}}, \bibnamefont{and}
  \bibinfo{author}{\bibfnamefont{H.}~\bibnamefont{Zatzkis}},
  \bibinfo{journal}{Phys. Rev.} \textbf{\bibinfo{volume}{80}},
  \bibinfo{pages}{81} (\bibinfo{year}{1950}).

\bibitem[{\citenamefont{Dirac}(1958)}]{dirac:58}
\bibinfo{author}{\bibfnamefont{P.~A.~M.} \bibnamefont{Dirac}},
  \bibinfo{journal}{Proc. Roy. Soc. A} \textbf{\bibinfo{volume}{246}},
  \bibinfo{pages}{333} (\bibinfo{year}{1958}).

\bibitem[{\citenamefont{Arnowitt et~al.}(1962)\citenamefont{Arnowitt, Deser,
  and Misner}}]{arnowitt:62}
\bibinfo{author}{\bibfnamefont{R.}~\bibnamefont{Arnowitt}},
  \bibinfo{author}{\bibfnamefont{S.}~\bibnamefont{Deser}}, \bibnamefont{and}
  \bibinfo{author}{\bibfnamefont{C.}~\bibnamefont{Misner}}, in
  \emph{\bibinfo{booktitle}{{Gravitation: An Introduction to Current
  Research}}}, edited by
  \bibinfo{editor}{\bibfnamefont{L.}~\bibnamefont{Witten}}
  (\bibinfo{publisher}{Whiley}, \bibinfo{address}{New York},
  \bibinfo{year}{1962}).

\bibitem[{\citenamefont{Dirac}(2001)}]{dirac:74}
\bibinfo{author}{\bibfnamefont{P.~A.~M.} \bibnamefont{Dirac}},
  \emph{\bibinfo{title}{{Lectures on Quantum Mechanics}}}
  (\bibinfo{publisher}{Dover Publications}, \bibinfo{year}{2001}).

\bibitem[{\citenamefont{Wheeler}(1968)}]{wheeler:68}
\bibinfo{author}{\bibfnamefont{J.~A.} \bibnamefont{Wheeler}}, in
  \emph{\bibinfo{booktitle}{{Battelles Rencontres: 1967 Lectures in Mathematics
  and Physics}}}, edited by \bibinfo{editor}{\bibfnamefont{C.~M.}
  \bibnamefont{DeWitt}} \bibnamefont{and} \bibinfo{editor}{\bibfnamefont{J.~A.}
  \bibnamefont{Wheeler}} (\bibinfo{publisher}{Benjamin}, \bibinfo{address}{New
  York}, \bibinfo{year}{1968}).

\bibitem[{\citenamefont{DeWitt}(1967)}]{dewitt:69}
\bibinfo{author}{\bibfnamefont{B.~S.} \bibnamefont{DeWitt}},
  \bibinfo{journal}{Phys. Rev.} \textbf{\bibinfo{volume}{160}},
  \bibinfo{pages}{1113} (\bibinfo{year}{1967}).

\bibitem[{\citenamefont{Kucha$\check{\textrm{r}}$}(1972)}]{kuchar:72}
\bibinfo{author}{\bibfnamefont{K.~V.} \bibnamefont{Kucha$\check{\textrm{r}}$}},
  \bibinfo{journal}{JMP} \textbf{\bibinfo{volume}{13}}, \bibinfo{pages}{768}
  (\bibinfo{year}{1972}).

\bibitem[{\citenamefont{Kucha$\check{\textrm{r}}$}(1973)}]{kuchar:73}
\bibinfo{author}{\bibfnamefont{K.~V.} \bibnamefont{Kucha$\check{\textrm{r}}$}},
  in \emph{\bibinfo{booktitle}{{Relativity, Astrophysics and Cosmology}}},
  edited by \bibinfo{editor}{\bibfnamefont{W.}~\bibnamefont{Israel}}
  (\bibinfo{publisher}{D. Reidel Publishing Company},
  \bibinfo{address}{Dordrecht-Holland, Boston-U.S.A.}, \bibinfo{year}{1973}),
  pp. \bibinfo{pages}{237--288}.

\bibitem[{\citenamefont{Tomonaga}(1946)}]{tomonaga:46}
\bibinfo{author}{\bibfnamefont{S.}~\bibnamefont{Tomonaga}},
  \bibinfo{journal}{PTP} \textbf{\bibinfo{volume}{I}}, \bibinfo{pages}{27}
  (\bibinfo{year}{1946}).

\bibitem[{\citenamefont{Schwinger}(1948)}]{schwinger:48}
\bibinfo{author}{\bibfnamefont{J.}~\bibnamefont{Schwinger}},
  \bibinfo{journal}{Phys. Rev.} \textbf{\bibinfo{volume}{74}},
  \bibinfo{pages}{1439} (\bibinfo{year}{1948}).

\bibitem[{\citenamefont{Ashtekar}(1986)}]{ashtekar:86}
\bibinfo{author}{\bibfnamefont{A.}~\bibnamefont{Ashtekar}},
  \bibinfo{journal}{Phys. Rev. Lett.} \textbf{\bibinfo{volume}{57}},
  \bibinfo{pages}{2244} (\bibinfo{year}{1986}).

\bibitem[{\citenamefont{Sen}(1982)}]{sen:82}
\bibinfo{author}{\bibfnamefont{A.}~\bibnamefont{Sen}}, \bibinfo{journal}{Phys.
  Lett. B} \textbf{\bibinfo{volume}{119}}, \bibinfo{pages}{89}
  (\bibinfo{year}{1982}).

\bibitem[{\citenamefont{Isham}(1993)}]{isham:93}
\bibinfo{author}{\bibfnamefont{C.}~\bibnamefont{Isham}}, in
  \emph{\bibinfo{booktitle}{{Integrable Systems, Quantum Groups and Quantum
  Field Theory}}}, edited by \bibinfo{editor}{\bibfnamefont{L.~A.}
  \bibnamefont{Ibort}} \bibnamefont{and} \bibinfo{editor}{\bibfnamefont{M.~A.}
  \bibnamefont{Rodr\'iguez}} (\bibinfo{publisher}{Kluwer Academic},
  \bibinfo{address}{Dordrecht, Boston}, \bibinfo{year}{1993}), pp.
  \bibinfo{pages}{157--287}.

\bibitem[{\citenamefont{Kucha$\check{\textrm{r}}$}(1992)}]{kuchar:92}
\bibinfo{author}{\bibfnamefont{K.~V.} \bibnamefont{Kucha$\check{\textrm{r}}$}},
  in \emph{\bibinfo{booktitle}{{Proceedings of the 4$^{th}$ Canadian Conference
  on General Relativity and Relativistic Astrophysics}}}, edited by
  \bibinfo{editor}{\bibfnamefont{G.}~\bibnamefont{Kunstatter}},
  \bibinfo{editor}{\bibfnamefont{D.}~\bibnamefont{Vincent}}, \bibnamefont{and}
  \bibinfo{editor}{\bibfnamefont{J.}~\bibnamefont{Williams}}
  (\bibinfo{publisher}{World Scientific}, \bibinfo{address}{Singapore},
  \bibinfo{year}{1992}), pp. \bibinfo{pages}{211--314}.

\bibitem[{\citenamefont{Kiefer}(2007)}]{kiefer:07}
\bibinfo{author}{\bibfnamefont{C.}~\bibnamefont{Kiefer}},
  \emph{\bibinfo{title}{{Quantum Gravity}}} (\bibinfo{publisher}{Oxford
  University Press}, \bibinfo{address}{New York}, \bibinfo{year}{2007}),
  \bibinfo{edition}{2nd} ed.

\bibitem[{\citenamefont{Torre}(1992)}]{torre:92}
\bibinfo{author}{\bibfnamefont{C.~G.} \bibnamefont{Torre}},
  \bibinfo{journal}{Phys. Rev. D} \textbf{\bibinfo{volume}{46}},
  \bibinfo{pages}{R3231} (\bibinfo{year}{1992}).

\bibitem[{\citenamefont{Ashtekar
  et~al.}(2006{\natexlab{a}})\citenamefont{Ashtekar, Pawlowski, and
  Sing}}]{ashtekar:06a}
\bibinfo{author}{\bibfnamefont{A.}~\bibnamefont{Ashtekar}},
  \bibinfo{author}{\bibfnamefont{T.}~\bibnamefont{Pawlowski}},
  \bibnamefont{and} \bibinfo{author}{\bibfnamefont{P.}~\bibnamefont{Sing}},
  \bibinfo{journal}{Phys. Rev. Lett.} \textbf{\bibinfo{volume}{96}}
  (\bibinfo{year}{2006}{\natexlab{a}}), \bibinfo{note}{arXiv:gr-qc/0602086v2}.

\bibitem[{\citenamefont{Ashtekar
  et~al.}(2006{\natexlab{b}})\citenamefont{Ashtekar, Pawlowski, and
  Sing}}]{ashtekar:06b}
\bibinfo{author}{\bibfnamefont{A.}~\bibnamefont{Ashtekar}},
  \bibinfo{author}{\bibfnamefont{T.}~\bibnamefont{Pawlowski}},
  \bibnamefont{and} \bibinfo{author}{\bibfnamefont{P.}~\bibnamefont{Sing}},
  \bibinfo{journal}{Phys. Rev. D} \textbf{\bibinfo{volume}{73}}
  (\bibinfo{year}{2006}{\natexlab{b}}), \bibinfo{note}{arXiv:gr-qc/0604013v3}.

\bibitem[{\citenamefont{Ashtekar
  et~al.}(2006{\natexlab{c}})\citenamefont{Ashtekar, Pawlowski, and
  Sing}}]{ashtekar:06c}
\bibinfo{author}{\bibfnamefont{A.}~\bibnamefont{Ashtekar}},
  \bibinfo{author}{\bibfnamefont{T.}~\bibnamefont{Pawlowski}},
  \bibnamefont{and} \bibinfo{author}{\bibfnamefont{P.}~\bibnamefont{Sing}},
  \bibinfo{journal}{Phys. Rev. D} \textbf{\bibinfo{volume}{74}}
  (\bibinfo{year}{2006}{\natexlab{c}}), \bibinfo{note}{arXiv:gr-qc/0607039v2}.

\bibitem[{\citenamefont{Ashtekar et~al.}(2007)\citenamefont{Ashtekar,
  Pawlowski, Sing, and Vandersloot}}]{ashtekar:07}
\bibinfo{author}{\bibfnamefont{A.}~\bibnamefont{Ashtekar}},
  \bibinfo{author}{\bibfnamefont{T.}~\bibnamefont{Pawlowski}},
  \bibinfo{author}{\bibfnamefont{P.}~\bibnamefont{Sing}}, \bibnamefont{and}
  \bibinfo{author}{\bibfnamefont{K.}~\bibnamefont{Vandersloot}},
  \bibinfo{journal}{Phys. Rev. D} \textbf{\bibinfo{volume}{75}}
  (\bibinfo{year}{2007}), \bibinfo{note}{arXiv:gr-qc/0612104v2}.

\bibitem[{\citenamefont{Husain and Winkler}(2004)}]{husain:03}
\bibinfo{author}{\bibfnamefont{V.}~\bibnamefont{Husain}} \bibnamefont{and}
  \bibinfo{author}{\bibfnamefont{O.}~\bibnamefont{Winkler}},
  \bibinfo{journal}{Phys. Rev. D} \textbf{\bibinfo{volume}{69}}
  (\bibinfo{year}{2004}), \bibinfo{note}{arXiv:gr-qc/0312094v1}.

\bibitem[{\citenamefont{Husain and Winkler}(2005)}]{husain:08}
\bibinfo{author}{\bibfnamefont{V.}~\bibnamefont{Husain}} \bibnamefont{and}
  \bibinfo{author}{\bibfnamefont{O.}~\bibnamefont{Winkler}},
  \bibinfo{journal}{Class. Quantum Grav.} \textbf{\bibinfo{volume}{22}},
  \bibinfo{pages}{L127} (\bibinfo{year}{2005}),
  \bibinfo{note}{arXiv:gr-qc/0410125v3}.

\bibitem[{\citenamefont{Ashtekar and Bojowald}(2006)}]{ashtekar:05}
\bibinfo{author}{\bibfnamefont{A.}~\bibnamefont{Ashtekar}} \bibnamefont{and}
  \bibinfo{author}{\bibfnamefont{M.}~\bibnamefont{Bojowald}},
  \bibinfo{journal}{Class. Quantum Grav.} \textbf{\bibinfo{volume}{23}},
  \bibinfo{pages}{391} (\bibinfo{year}{2006}),
  \bibinfo{note}{arXiv:gr-qc/0509075v1}.

\bibitem[{\citenamefont{Modesto}(2004)}]{modesto:04}
\bibinfo{author}{\bibfnamefont{L.}~\bibnamefont{Modesto}},
  \bibinfo{journal}{Phys. Rev. D} \textbf{\bibinfo{volume}{70}},
  \bibinfo{pages}{124009} (\bibinfo{year}{2004}),
  \bibinfo{note}{arXiv:gr-qc/0407097v2}.

\bibitem[{\citenamefont{Modesto}(2006)}]{modesto:06}
\bibinfo{author}{\bibfnamefont{L.}~\bibnamefont{Modesto}},
  \bibinfo{journal}{Class. Quantum Grav.} \textbf{\bibinfo{volume}{23}},
  \bibinfo{pages}{5587} (\bibinfo{year}{2006}),
  \bibinfo{note}{arXiv:gr-qc/0509078v2}.

\bibitem[{\citenamefont{Stephani et~al.}(2003)\citenamefont{Stephani, Kramer,
  MacCallum, Hoenselaers, and Herlt}}]{stephani:03}
\bibinfo{author}{\bibfnamefont{H.}~\bibnamefont{Stephani}},
  \bibinfo{author}{\bibfnamefont{D.}~\bibnamefont{Kramer}},
  \bibinfo{author}{\bibfnamefont{M.}~\bibnamefont{MacCallum}},
  \bibinfo{author}{\bibfnamefont{C.}~\bibnamefont{Hoenselaers}},
  \bibnamefont{and} \bibinfo{author}{\bibfnamefont{E.}~\bibnamefont{Herlt}},
  \emph{\bibinfo{title}{{Exact Solutions of Einstein's Field Equations}}}
  (\bibinfo{publisher}{Cambridge University Press},
  \bibinfo{address}{Cambridge, United Kindom}, \bibinfo{year}{2003}),
  \bibinfo{edition}{2nd} ed.

\bibitem[{\citenamefont{Blyth and Isham}(1975)}]{blyth:75}
\bibinfo{author}{\bibfnamefont{W.~F.} \bibnamefont{Blyth}} \bibnamefont{and}
  \bibinfo{author}{\bibfnamefont{C.~J.} \bibnamefont{Isham}},
  \bibinfo{journal}{Phys. Rev. D} \textbf{\bibinfo{volume}{11}},
  \bibinfo{pages}{768} (\bibinfo{year}{1975}).

\bibitem[{\citenamefont{Abramowitz and Stegun}(1972)}]{abramowitz:72}
\bibinfo{editor}{\bibfnamefont{M.}~\bibnamefont{Abramowitz}} \bibnamefont{and}
  \bibinfo{editor}{\bibfnamefont{A.}~\bibnamefont{Stegun}}, eds.,
  \emph{\bibinfo{title}{{Handbook of Mathematical Functions with Formulas,
  Graphs and Mathematical Tables}}} (\bibinfo{publisher}{Dover},
  \bibinfo{address}{New York}, \bibinfo{year}{1972}).

\bibitem[{\citenamefont{Gradshteyn and Ryzhik}(2007)}]{gradshteyn:07}
\bibinfo{author}{\bibfnamefont{I.~S.} \bibnamefont{Gradshteyn}}
  \bibnamefont{and} \bibinfo{author}{\bibfnamefont{I.~M.}
  \bibnamefont{Ryzhik}}, \emph{\bibinfo{title}{{Table of Integrals, Series, and
  Products}}} (\bibinfo{publisher}{Elsevier Academic Press},
  \bibinfo{address}{San Diego, London}, \bibinfo{year}{2007}),
  \bibinfo{edition}{7th} ed.

\bibitem[{\citenamefont{Corichi et~al.}(2007)\citenamefont{Corichi,
  Vuka$\check{\textrm{s}}$inac, and Zapata}}]{corichi:07}
\bibinfo{author}{\bibfnamefont{A.}~\bibnamefont{Corichi}},
  \bibinfo{author}{\bibfnamefont{T.}~\bibnamefont{Vuka$\check{\textrm{s}}$inac%
}}, \bibnamefont{and} \bibinfo{author}{\bibfnamefont{J.~A.}
  \bibnamefont{Zapata}}, \bibinfo{journal}{Phys. Rev. D}
  \textbf{\bibinfo{volume}{76}} (\bibinfo{year}{2007}).

\bibitem[{\citenamefont{Husain}(1987)}]{husain:87}
\bibinfo{author}{\bibfnamefont{V.}~\bibnamefont{Husain}},
  \bibinfo{journal}{Class. Quantum Grav.} \textbf{\bibinfo{volume}{4}},
  \bibinfo{pages}{1587} (\bibinfo{year}{1987}).

\bibitem[{\citenamefont{Kiefer}(1993)}]{kiefer:93}
\bibinfo{author}{\bibfnamefont{C.}~\bibnamefont{Kiefer}},
  \emph{\bibinfo{title}{{The Semiclassical Approximation to Quantum Gravity}}}
  (\bibinfo{year}{1993}), \bibinfo{note}{arXiv:gr-qc/9312015v1}.

\end{thebibliography}

\end{document}